\documentclass[12pt,a4paper]{article}

\usepackage[utf8]{inputenc}
\usepackage{cite}
\usepackage{hyperref}
\usepackage{setspace}
\hypersetup{
	colorlinks=true,
	linkcolor=dark-blue,
	citecolor=dark-red,
	urlcolor=dark-green,
	linktoc=page
}
\usepackage{color}
\definecolor{dark-gray}{gray}{0.20}
\definecolor{gray}{gray}{0.30}
\definecolor{light-gray}{gray}{0.80}
\definecolor{dark-red}{rgb}{0.7,0,0}
\definecolor{dark-green}{rgb}{0.1,0.4,0}
\definecolor{dark-blue}{rgb}{0.3,0.3,0.7}
\definecolor{light-blue}{rgb}{0.8,0.8,1}
  \usepackage{a4wide}
  \usepackage{latexsym}
  \usepackage{epsf}
  \usepackage{bbm}
  \usepackage{graphicx}
  \usepackage{amsmath,amssymb,amsthm}
\usepackage{verbatim}

\renewcommand{\d}{\textrm{d}}
\newcommand{\sgn}{\textrm{sgn}\,}

\newcommand{\dd}{\mathrm{d}}
\newcommand{\f}[2]{\frac{#1}{#2}}
\newcommand{\e}{\textrm{e}}

\newcommand{\vol}{\text{vol}}

\pdfoutput=1							

\csname @addtoreset\endcsname{equation}{section}

\usepackage{tikz}
\usetikzlibrary{arrows,automata,positioning,calc,trees,decorations.pathmorphing,decorations.markings,3d,fadings}
\tikzset{snake it/.style={decorate, decoration={snake,segment length=3pt, amplitude=1pt}}}
\usepackage{tikz}
\usetikzlibrary{positioning,arrows}
\usetikzlibrary{decorations.pathmorphing}
\usetikzlibrary{decorations.markings}
\usetikzlibrary{decorations.text}
\usetikzlibrary{calc}

\usetikzlibrary{arrows,automata,positioning,calc,trees,decorations.pathmorphing,decorations.markings,3d}
\usepackage{pgfplots}

\usepackage{tikz}

\usetikzlibrary{decorations.markings}

\begin{document}

\begin{flushright}
\tt\small{IFT-UAM/CSIC-19-95}\\
\tt\small{ROM2F/2019/05}
\end{flushright}
\vspace{1.2cm}

\begin{center}

{\LARGE {\bf Anti-brane singularities as red herrings }}  \\

\vspace{1.5 cm} {\large J. Bl{\aa}b{\"a}ck$^{a}$, F. F Gautason$^{b}$, A. Ruip\'erez$^{bc}$, T. Van
Riet$^b$ }\\
\vspace{0.5 cm}  \vspace{.15 cm} {${}^a$ Dipartimento di Fisica, Universit\`a di Roma ``Tor Vergata'' $\&$ \\INFN - Sezione di Roma2
Via della Ricerca Scientifica 1, 00133 Roma, Italy\\
${}^b$ Instituut voor Theoretische Fysica, KU Leuven,\\
Celestijnenlaan 200D B-3001 Leuven, Belgium\\
${}^c$ Instituto de F\'isica Te\'orica UAM/CSIC\\
C/ Nicol\'as Cabrera, 13–15, C.U. Cantoblanco, E-28049 Madrid, Spain}

\vspace{0.7cm} {\small \upshape\ttfamily \href{mailto:johan.blaback@roma2.infn.it}{johan.blaback@roma2.infn.it}, \href{mailto:alejandro.ruiperez@uam.es}{alejandro.ruiperez@uam.es}, \\ \{\href{mailto:ffg@kuleuven.be}{ffg}, \href{mailto:thomas.vanriet@kuleuven.be}{thomas.vanriet}\} \textcolor{dark-green}{@kuleuven.be}
 }  \\

\vspace{2cm}

{\bf Abstract}
\end{center}

{\small Unphysical 3-form flux singularities near anti-branes have been argued to get resolved in the classical supergravity regime when brane polarisation is properly taken into account. The only example that does not seem to fit this logic is the $\overline{\text{D6}}$-brane because of a no-go theorem for well behaved supergravity solutions with negative D6 charge. In this paper we first review the existing results demonstrating how brane polarisation resolves singularities for $\overline{\text{D3}}$-branes and then we improve on the description of the polarisation of $\overline{\text{D6}}$-branes into KK5 dipoles. We argue that the meta-stable state carries exactly zero (anti-)D6 charge, which is the unique way around the no-go theorem.  We then  provide numerical evidence for well-behaved solutions that describe such meta-stable states.
}

\newpage

\section{Introduction}
Branes and anti-branes preserve different supercharges. When both objects are residing in a single spacetime the setup therefore  tends to break all supersymmetries. A related phenomenon is the non-zero force that attracts branes and anti-branes eventually leading to annihilation into closed string radiation. For that reason anti-branes are not the best candidate ingredients when the goal is to break supersymmetry in a meta-stable fashion. It is however possible to dissolve branes into fluxes and then combine with an anti-brane. Fluxes can carry brane charge (dissolved branes) due to the transgression terms in the Bianchi identity for Ramond-Ramond (RR) field strengths:
\begin{equation}
\d F_{8-k} = H_3\wedge F_{6-k}+Q\delta_{9-k}\,.
\end{equation}
Here $H_3$ is the NSNS 3-form flux, $F_{8-k}$ and $F_{6-k}$ are RR fluxes and $Q\delta_{9-k}$ represents the localised magnetic RR charges induced by e.g.~(anti-)D$k$-branes. Note that $Q\delta_{9-k}$
is a $(9-k)$ form distribution. If the orientation of the latter form is opposite to $H_3\wedge F_{6-k}$ we denote the source \emph{anti-}brane. These still break supersymmetry but it is not as clear that combining fluxes and anti-branes leads to a perturbative decay. The reason is that brane-flux decay has to proceed via the nucleation of branes out of the flux cloud, which can then annihilate the anti-brane. It is reasonable to expect that nucleation of branes out of flux clouds to be a non-perturbative process and hence anti-branes in flux clouds of opposite charge might be the ideal setup for meta-stable supersymmetry breaking in various contexts ranging from holographic duals to dynamical supersymmetry-breaking in strongly coupled gauge theories \cite{Maldacena:2001pb, Kachru:2002gs}, to non-supersymmetric black hole microstates \cite{Bena:2011fc} and cosmological model building \cite{Kachru:2003aw, Kachru:2003sx}.

Flux solutions in supergravity feature non-trivial topological cycles which are threaded by fluxes of various field strengths available in the theory. Let $K$ be the flux quanta of the NSNS 3-form along one 3-cycle (commonly refereed to as the \emph{B}-cycle) and $M$ the quantum of RR $F_{6-k}$ form along some $(6-k)$-cycle (a.k.a.~A-cycle). When $p$ $\overline{\text{D}k}$-branes are introduced into the geometry which are point-like on the two cycles mentioned, then brane-flux decay occurs when fluxes of either $H_3$ or $F_{6-k}$ drop by one unit at the expense of producing either $M$ or $K$ D$k$-branes. After decay this results in $M-p$ or $K-p$ D$k$-branes (if $p<K$ and $p<M$).

For the case of $\overline{\text{D3}}$-branes inside the Klebanov-Strassler throat solution \cite{Klebanov:2000hb}, Kachru, Pearson, and Verlinde (KPV) \cite{Kachru:2002gs} have described rather explicitly how this brane-flux decay proceeds in the NSNS sector and computed the effective potential for this decay process. The result is that for $p/M$ small enough the decay is indeed non-perturbative and one can obtain a long lived vacuum. A priori it is not obvious what an effective potential in this case would mean since there is  no obvious continuous degree of freedom (field) that mediates the decay. This issue was clarified by KPV in \cite{Kachru:2002gs} by showing that brane-flux decay proceeds via brane polarisation {\`a} la Myers \cite{Myers:1999ps}. The $p$ $\overline{\text{D3}}$-branes ``puff'' into a spherical NS5-brane with worldvolume fluxes. That spherical NS5-brane wraps a contractible 2-cycle inside the 3-cycle threaded by $F_3$ flux. The brane feels two forces; on the one hand it would tend to collapse under its own weight and, on the other hand, it is forced to expand because of the background NSNS 3-form flux. KPV showed that at small enough $p/M$ the whole setup finds a local balance of forces. The continuous degree of freedom that mediates the decay is the NS5 position on the $S^3$.

A similar story holds for the RR channel and for $\overline{\text{D}k}$-branes with $k\neq 3$ as long as $k<7$. A pedagogical treatment of such examples of brane flux decay in simple flux throats for $\overline{\text{D}k}$-branes with general $k$ is given in reference \cite{Gautason:2015tla}.

Despite the elegant picture, the arguments behind meta-stable states for small $p/M$ have been questioned over the years. The main issue is that none of the probe computations were carried in a regime were they are supposed to work. For example the NS5 action used in \cite{Kachru:2002gs} is not a valid description at small string coupling and neither is the non-Abelian D3 action that was used in the same paper. This is not particular to $\overline{\text{D3}}$'s but applies to all situations including $\overline{\text{M2}}$-brane meta-stable states \cite{Klebanov:2010qs}. One always has to rely on a description using a duality but in the ``wrong regime". Not unrelated are worries about the backreaction of the probe. These worries cumulated in a series of papers \cite{Bena:2010gs, Giecold:2011gw, Bena:2011wh, Bena:2012ek, Bena:2012tx, Bena:2012vz, Bena:2013hr, Blaback:2011nz, Blaback:2011pn, Blaback:2013hqa, Cottrell:2013asa, Blaback:2014tfa, Gautason:2013zw, Giecold:2013pza, Hartnett:2015oda} following on \cite{Bena:2009xk} (and \cite{McGuirk:2009xx}), which claimed that the supergravity solutions corresponding to anti-branes have diverging NSNS 3-form fluxes
\begin{equation}
e^{-\phi}|H_3|^2\rightarrow \infty\,,
\end{equation}
near the anti-branes. It was argued that the singularity cannot be interpreted as the local $H_3$ singular flux sourced by the polarised NS5-brane because it has the wrong degree of divergence and points in the wrong directions.

Singular solutions of supergravity should always be treated with care since strictly speaking they bring us out of the supergravity regime and they could signal inconsistencies. Indeed, the singular flux can be interpreted as an indication of a perturbative runaway  \cite{Blaback:2012nf, Danielsson:2014yga}. Once the background would be non-stationary the singularity would get resolved: As time goes by the fluxes are being drawn towards the anti-branes until they reach a critical value at which perturbative brane-flux decay sets in, naturally resolving the infinite flux by letting the branes decay away. These explanations of the singularity were recently argued to be erroneous in \cite{Michel:2014lva, Cohen-Maldonado:2015lyb, Cohen-Maldonado:2016cjh} (see also \cite{Polchinski:2015bea}).

The arguments of \cite{Michel:2014lva} relied on the situation with a single anti-brane ($p=1$) and is therefore not amendable to a supergravity analysis which requires $g_sp\gg 1$. Instead the arguments were based on brane effective field theory methods and string theory intuition. The main result was that the flux clumping should be cut-off at string scale distance and then the clumping is still sufficiently small in order to prevent perturbative decay.

Instead the computations carried out in \cite{Cohen-Maldonado:2015ssa, Cohen-Maldonado:2016cjh} are applicable in the supergravity regime and showed that the singularities were an artefact of using a too restrictive supergravity Ansatz; in short the Ansatz did not include the polarised brane, which is predicted to be there from the probe analysis. This was already suggested earlier by Dymarsky in \cite{Dymarsky:2011pm} (see also \cite{Bena:2014bxa, Bena:2014jaa} for related comments). Hence the singularities were indeed signalling instabilities, but not dangerous ones, they signalled that the $\overline{\text{D3}}$ wants to polarise.

The same reasoning works for $\overline{\text{M2}}$-branes and all $\overline{\text{D}k}$-branes with $k<5$ \cite{Cohen-Maldonado:2016cjh}. The cases of an $\overline{\text{D5}}$ and $\overline{\text{D6}}$ are however special. The $\overline{\text{D5}}$ does not decay to a single spherical NS5-branes but rather a pair of NS5-branes which between them carry D5-brane charge. For D6-branes the situation is even more confusing. Brane-flux decay in the RR channel proceeds via D8-branes that discharge the $F_0$ flux but it can be shown \cite{Gautason:2015ola} not to occur in the flux throat solution of \cite{Janssen:1999sa}. Hence, as could be expected from naive application of T-duality, the decay proceeds in the NSNS channel via the polarisation into KK5 monopoles. This process is not as clear conceptually as the other polarisation channels but can be dealt with by considering the aforementioned T-dual process of $\overline{\text{D5}}$-brane polarising into a pair of NS5-branes \cite{Gautason:2015tla, Danielsson:2016cit}. Indeed for $\overline{\text{D5}}$-branes polarising to a pair of D5-branes the A-cycle is a circle on which T-duality can be performed.

We recall the details of these special cases below. An important issue is that, unlike $\overline{\text{D3}}$-branes, $\overline{\text{D5}}$ and $\overline{\text{D6}}$-branes polarise into branes of the same codimension. Hence they ``attract" the background $H_3$-flux more intensively than in the $\overline{\text{D3}}$ case.\footnote{Recall that KK5 monopoles in supergravity assume a circular isometry along the KK direction making them effectively 6-branes.} Furthermore the argument of \cite{Cohen-Maldonado:2015ssa, Cohen-Maldonado:2016cjh} outlined above fails for the case of the $\overline{\text{D6}}$-branes.\footnote{The reason is that flux backgrounds carrying D6-brane charge does not have a regular asymptotic regime (UV) as explained further in this paper.} The supergravity analysis of $\overline{\text{D6}}$ background is however the case where most analytic control is at hand. This is a consequence of the isometry preserved by the brane. In fact it is a simple task to show that whenever $Q_6<0$ the $H_3$-flux density  is singular \cite{Blaback:2011nz, Blaback:2011pn}. In this paper we show that despite this argument, there is no reason to expect a singularity in the fully backreacted solutions. The reason is surprisingly simple; the meta-stable probes polarise rather strongly until they reach exactly the point where the monopole $\overline{\text{D6}}$-brane charge vanishes. This simple observation was not made in \cite{Danielsson:2016cit}. The computation carried out in \cite{Danielsson:2016cit} assumed that the monopole charge was still non-zero and negative and then performed a cut-off at string scale (as done in \cite{Michel:2014lva}) to compute the corrected probe potential finding it to be runaway.\footnote{This situation is different from meta-stable $\overline{\text{D3}}$ states where the critical points of vanishing monopole charge exactly occur at the threshold of loosing the meta-stable state: $p/M=0.08$.} We conclude that anti-branes are not examples where probe computations clash with fully backreacted treatments. This does not mean that anti-brane supersymmetry-breaking is without possible caveats. See \cite{Bena:2014bxa, Bena:2014jaa, Bena:2015kia, Bena:2016fqp, Danielsson:2015eqa}.

The rest of this paper is organised as follows. In Section \ref{sec:singularD6} we recall the arguments of \cite{Blaback:2011nz} for why configurations with non-zero $\overline{\text{D6}}$ charges have unphysical singularities and why polarisation resolves the singularities for $k<5$. In Section \ref{sec:KK5} we recall and improve on the probe computations of \cite{Danielsson:2016cit} by incorporating the tachyon dynamics, which turns out crucial in order to demonstrate the existence of meta-stable probes. We emphasise that the meta-stable state carries no $\overline{\text{D6}}$ charge and that this is what is need to evade the no-go theorems for well behaved solutions. In Section \ref{sec:regular} we briefly speculate about the existence of regular solutions and give partial evidence for them numerically, and we end with a discussion in Section \ref{sec:discussion}.

\section{Singular/regular anti-branes }\label{sec:singularD6}

\subsection{Why polarisation cures singularities for D$(k<5)$-branes}

The reason polarisation resolves singularities for $\overline{\text{D}k}$-branes with $k<5$, $\overline{\text{M2}}$ and $\overline{\text{M5}}$-branes is surprisingly simple and analytic \cite{Cohen-Maldonado:2015ssa, Cohen-Maldonado:2016cjh}. Therefore we sketch it here for the case of $\overline{\text{D3}}$-branes.

The following Ansatz for the would-be supergravity solution describing an $\overline{\text{D3}}$-brane can be written as (in Einstein frame)
\begin{align}
& \d s^2_{10} = e^{2A}(-\e^{2f}(\dd x^0)^2 +(\dd x^1)^2+(\dd x^2)^2+(\dd x^3)^2) + \d s^2_6\,,\nonumber\\
&C_4 = \alpha\, \vol_4\,,\nonumber \\
& H_7 = \e^{-\phi}\star_{10} H_3 = -\vol_4\wedge \left[\alpha F_3 + \dd A_2\right]~.
\end{align}
where the dilaton $\phi$ and the functions $\alpha$ and $A$ depend only on the coordinates of the transverse metric $\d s^2_6$ that we leave unspecified. We have introduced the notation $\vol_4=\d x^0 \wedge \d x^1\wedge \d x^2\wedge \d x^3$ and the 2-form $A_2$ which is a gauge potential that determines $H_7$. A priori the 3-form $X_3 = \dd A_2$ could contain non-exact pieces but they are shown to vanish in \cite{Gautason:2013zw}. Up to a $C_4$ gauge transformation this is the most general Ansatz. For completeness we have included the possibility of having finite-temperature branes but we will mostly focus on the zero-temperature case.  Starting from this Ansatz it was shown in \cite{Cohen-Maldonado:2015ssa, Cohen-Maldonado:2016cjh} inspired by  \cite{Gautason:2013zw,Blaback:2014tfa} that any consistent background must obey a generalised Smarr relation
\begin{equation}\label{smarr}
{\cal E} = \f{5}{4}\f{\kappa{\cal A}}{8\pi G_N} + \Phi_\text{D3} Q_\text{D3} + \Phi_\text{NS5} Q_\text{NS5}\,,
\end{equation}
which relates quantities measured in the UV and the IR. More precisely the energy density ${\cal E}$, the surface gravity $\kappa$ (which is related to temperature), and the D3-brane charge $Q_{\text{D3}}$ are all defined in the asymptotic region of space-time, or the UV. On the other hand the generalised area ${\cal A}$ (related to entropy density) and chemical potentials $\Phi_{\text{D3},\text{NS5}}$ are defined at the horizon of the brane system.   In particular
\begin{equation}
\Phi_\text{D3} = \alpha|_\text{horizon}~,
\end{equation}
and $\Phi_\text{NS5}$ is determined by $A_2$ on the horizon. The NS5-brane charge $Q_\text{NS5}$ is also defined in the near-horizon region since it denotes the NS5 dipole charge (which is proportional to number of NS$5$-branes) of a polarised state. We refer to \cite{Cohen-Maldonado:2015ssa, Cohen-Maldonado:2016cjh} for details on how these quantities are defined. The power of the Smarr relation (\ref{smarr}) for our purposes, is the ability to translate regularity conditions which are easy to determine in the near-horizon (or IR) region to the  asymptotic (or UV) region.
The central quantity in the discussion that follows will be the $H_3$ energy density
\begin{equation}\label{fluxdens}
e^{-\phi}|H_3|^2 = e^{\phi-8A}|\alpha F_3+\dd A_2|^2\,,
\end{equation}
together with the Smarr relation \eqref{smarr}. Let us focus on the case of $\overline{\text{D3}}$-branes at zero temperature such that the first term in \eqref{smarr} vanishes. Using \eqref{fluxdens} and the standard behaviour of $A$ and $\phi$ close to D3-branes, one can show \cite{Cohen-Maldonado:2015ssa} that $\Phi_\text{D3}$ must vanish for any solution that either has no singularity in $H_3$ flux or one that can be interpreted as sourced by NS5-branes. For point-like or smeared branes one can show that the last term in \eqref{smarr} vanishes. This is because the horizon topology of the brane system must be non-trivial in order for $Q_{\text{NS5}}$ to be non-zero. This is the same as saying that $\overline{\text{D3}}$-branes must polarise in order to carry NS5 dipole charge. Finally recall that $\overline{\text{D3}}$-branes, which break supersymmetry, have ${\cal E}>0$. In fact their energy is simply two times the warped down D3-tension. Combining these facts we see that the Smarr relation implies that the chemical potential  $\Phi_\text{D3}$ is non-zero. We therefore arrive at a contradiction. Furthermore by  taking $\Phi_\text{D3}$ non-zero one reproduces the singularities found in \cite{Bena:2009xk} showing the consistency of our approach. A slightly more rigorous approach is to introduce a small non-zero temperature for which we expect a regular horizon surrounding the system of $\overline{\text{D3}}$-branes. As explained in \cite{Cohen-Maldonado:2016cjh}, the Raychaudhuri equation combined with the Einstein equation implies that for a regular horizon $H_7$ must vanish when restricted to it.  If it does not vanish one can show that the flux density of $H_3$ \eqref{fluxdens} diverges as before.  The zero-temperature limit therefore reproduces our previous result. The only way to maintain both ${\cal E}>0$ and $\Phi_\text{D3}=0$ is to have a non-zero $Q_5$. In other words, the Ansatz should be rich enough to allow for the NS5-brane, predicted in the probe analysis, to appear and backreact on the geometry.

\subsection{Singular $\overline{\text{D6}}$-branes}

To describe $\overline{\text{D6}}$-branes we first need a flux throat in which the brane is immersed. This solution was found by Janssen et al.~in \cite{Janssen:1999sa} and we briefly recall it here. The solution in string frame is given by\footnote{From now on we work in conventions where $\ell_s=1$.}
\begin{equation}\label{eq:fluxthroat1}
\begin{aligned}
\d s^2=&\,S^{-1/2}\d s_7^2+S^{1/2}\left(\d r^2+r^2\d\Omega_2^2\right)\,,\quad e^{\phi}=g_s\, S^{-3/4}\,, \\
F_2=&\,-g_s^{-1} S' r^2\Omega_2\,,\qquad\qquad\qquad\qquad~ H_3= - g_sM\, r^2\,\d r\wedge \Omega_2\,,
\end{aligned}
\end{equation}
where $F_0=M$, $\d{s}_7^2$ is the 7-dimensional Minkowski metric, $S$ is a function of the radial coordinate $r$, and $\Omega_2$ is the volume form on the 2-sphere. We denote $r$-derivatives of the function $S$ by a prime. The Bianchi identity of $F_2$ implies a second order differential equation for $S$ given by
\begin{equation}
S''+\f{2S'}{r}+g_s^2 M^2=0\,.
\end{equation}
The solution of which is
\begin{equation}\label{S-solution}
S=v^2-\frac{g_s^2M^2 r^2}{6} + \f{q}{r}\ ,
\end{equation}
with $v$ and $q$ two independent integration constants. Note that $q$ has the physical interpretation of localised D6-brane sources at the origin $r=0$ but since we are interested in a pure flux background we set it to zero. This is the flux background solution to which we will add probe $\overline{\text{D6}}$-branes. The fluxes themselves carry delocalised D6-brane charge throughout the throat. It is easy to verify that probe D6-branes feel no-forces in this solution whereas $\overline{\text{D6}}$-branes are pushed towards $r=0$. Because of the no-force property of the solution we often refer to it as being ``extremal'' or even ``BPS''. Note however that all supersymmetries are broken by the solution. The supersymmetric version of the above flux background are also known explicitly, but they break the Poincar{\'e} symmetries of the 7-dimensional metric \cite{Janssen:1999sa, Imamura:2001cr}. We note that the supersymmetry breaking featured in the flux background \eqref{eq:fluxthroat1} is nothing but a T-dual to the well-known supersymmetry breaking in IIB supergravity with warped Calabi-Yau geometries \cite{Giddings:2001yu} for which the ISD three-form fluxes have a non-vanishing $(0,3)$ piece \cite{Blaback:2012mu, Blaback:2013taa}. Although the flux background is regular at $r=0$ it possesses a curvature singularity at $r=r_0 =\sqrt{6} v/(g_sM)$. This singularity can be interpreted as the backreaction of a localised O6 plane which we discuss in some detail in Appendix \ref{app:O6}.

Let us now consider a more general Ansatz which could describe backreacting $\overline{\text{D6}}$-branes. This Ansatz was discussed in detail in \cite{Blaback:2011nz} and in Einstein frame is given by
\begin{equation}
\begin{split}
& \d s^2 =e^{2A(r)}\d s^2_7 + e^{2B(r)}[\d r^2 + r^2\d\Omega_2^2]\,,\label{Ansatz}\\
& F_2=-e^{-7A}\star_3\d\alpha\,,\qquad H_3= - \alpha e^{-7A+2\phi}\star_3 F_0\,,
\end{split}
\end{equation}
where $\star_3$ is the Hodge star including the conformal factor $\e^{B(r)}$. The useful aspect of $\overline{\text{D6}}$-branes is that their backreaction can be described using ODE's and the form of these ODE's was discussed and analysed in detail in \cite{Blaback:2011nz, Blaback:2011pn,Bena:2012tx,Junghans:2014wda}. The only aspects we need here are the following observations made in \cite{Blaback:2011nz}:
\begin{enumerate}
\item The equations of motion imply that at any regular point in the geometry the sign of $\alpha$ equals the sign of its second derivative $\alpha''$: $\sgn \alpha = \sgn \alpha''$.
\item The derivative of $\alpha$ near the source at $r=0$ is determined by its charge: $\sgn \alpha' = \sgn Q$.
\item If the anti-brane has confined backreaction effects then the solution should asymptotically approach the BPS solution (\ref{eq:fluxthroat1}). This implies that asymptotically $\alpha>0$.
\item Regular $H_3$ flux at the source near $r=0$ requires $\alpha(0)=0$.
\end{enumerate}
It is easy to show that these conditions cannot all be satisfied simultaneously if there is a source with negative D6 charge at the origin. It was shown in \cite{Bena:2012tx} that the $\overline{\text{D6}}$-branes do not polarise into D8-branes, which could have resolved the singularities as shown for supersymmetric $\overline{\text{D6}}$-branes in AdS$_7$ \cite{Apruzzi:2013yva, Junghans:2014wda}.\footnote{In AdS $\overline{\text{D6}}$-branes can be BPS.} So we are left with the NSNS channel that proceeds to the polarisation into KK5-branes. The would-be polarised state of KK5-branes with D6 charge which we discussed below should be captured by the same Ansatz if we assume that the KK-direction of the monopole is an isometry direction. And so the polarised state would again be plagued with the same singular behaviour. We are therefore seemingly left with the option that perhaps the singular $H_3$-flux is contained within a string length from the origin and should be dealt with using brane effective field theory. It was shown in \cite{Danielsson:2016cit} that for large $g_s p$ and $M$ the flux clumping extends a macroscopic distance away from the singularity causing perturbative brane-flux decay. However, as we show below this conclusion missed one crucial option which the probe computation reveals; the polarisation into KK5-branes creates a meta-stable state with exactly zero anti-brane charge $Q_6=0$. Let us emphasise that the meta-stable state is still there, but due to the interaction of the polarised KK5-brane with the surrounding flux cloud, the charge of the polarised state vanishes. In this case the no-go theorem reviewed above is evaded and there is no argument agains finding a well-behaved supergravity solution. We will provide numerical evidence that such solutions exist. Before doing that we first review and improve upon the probe computation of \cite{Danielsson:2016cit} for studying brane-flux decay of $\overline{\text{D6}}$.

\section{$\overline{\text{D6}}$ $\rightarrow$ KK5 polarisation}\label{sec:KK5}

\begin{figure}[ht!]
	\begin{center}
		\includegraphics{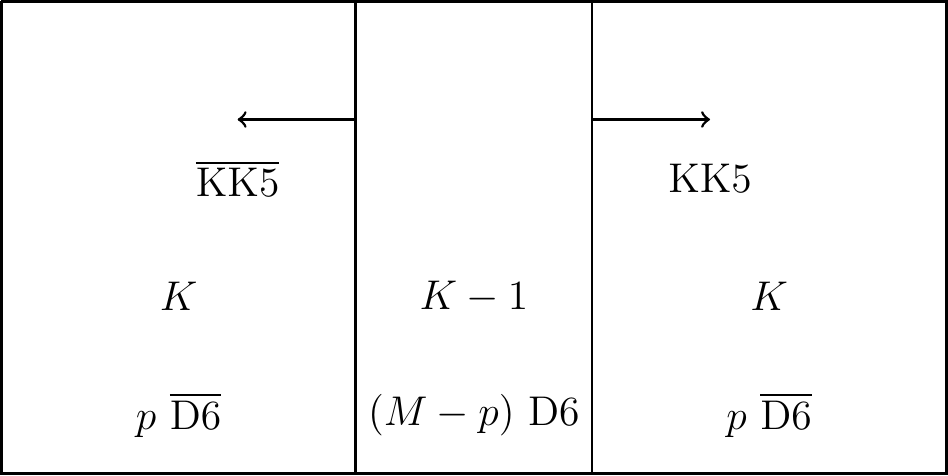}
		\caption{\it Decay of $\overline{\text{D6}}$-branes via brane polarisation into a pair of KK5-branes. The horizontal line correspond to the isometric direction $\psi$ along which the motion of the pair takes place. The vertical line represents the worldvolume directions.}
		\label{decay}
	\end{center}
\end{figure}

Let us now consider the insertion of $p$ \emph{probe} $\overline{\text{D6}}$-branes into the flux throat solution described in the previous section. We make sure that $p/M\ll 1$  so that their backreaction can  in principle be ignored. Some details of how $\overline{\text{D6}}$-branes (in the background under consideration) decay were explained in Ref.~\cite{Danielsson:2016cit} and here we will improve on that understanding.  Let us briefly recall the main idea. $\overline{\text{D6}}$-branes polarise into a pair of KK5/$\overline{\text{KK5}}$-branes, which can be understood as five-branes that are smeared over a circle inside the D6 flat worldvolume. This circle will be parametrised by the coordinate $\psi\sim \psi+1$ and then we will have that $ds_7^2=ds_6^2+d\psi^2$. As we will see in more detail later, the KK5 dipole carries an amount of D6-charge which depends on the relative distance between the brane and the anti-brane. Initially the charge is $-p$ when the branes are on top of each other and $M-p$ when they find each other again at the other side of the circle, as illustrated in Fig.~\ref{decay}. We want to know if the KK5 dipole finds a meta-stable state other than the true vacuum. To this aim, we will follow the approach of \cite{Kachru:2002gs} which basically consists of computing an effective potential for the degrees of freedom of the system, which are scalars parametrising the positions of the branes in the transverse directions, and check whether it has a minimum or not. Since these probe potentials are invariant under T-duality, we find it convenient to work in the T-dual frame, where D6 and KK5-branes are mapped to D5 and NS5-branes respectively.

\subsection{Effective potential for the NS5 dipole}

First of all, let us write the T-dual version of the background presented in (\ref{eq:fluxthroat1}), which is (in string frame)
\begin{eqnarray}
\d s^2&=&S^{-1/2}\dd s_6^2+S^{1/2}\left(\dd\psi^2+\dd r^2+r^2\dd\Omega_2^2\right)\ , \\
\label{eq:Tdualfluxthroat2}
F_1&=&M\,\dd\psi\ . \\
F_3&=&g_s^{-1}\tilde\star_4 \dd S\ , \\
e^{2\phi}&=& g_s^2 S^{-1} \\
H_3&=&g_s M\, r^2\,\dd r\wedge \Omega_2\, ,
\end{eqnarray}
where $\tilde\star_4$ is the Hodge dual with respect to the unwarped transverse metric and $S$ is the same function as in \eqref{S-solution}. $\overline{\text{D5}}$-probes inserted in this background feel a force that pushes them towards $r=0$. Taking this into account, it is clear that the degrees of freedom are only the scalars parametrising the position of the branes in the $\psi$ direction, that we denote by $\Psi_{\pm}$. The effective action describing the dynamics of the NS5/$\overline{\text{NS5}}$ pair would be the sum of three contributions
\begin{equation}
S_{\text{pair}}=S_{+}+S_{-}+S_{\text{int}}\ ,
\end{equation}
where \cite{Gautason:2015tla,Danielsson:2016cit}
\begin{equation}\label{eq:actionNS5}
S_{\pm}=-\mu_{5}\left\{\int_{W^6_\pm}\d^6x\,e^{-2\phi}\sqrt{-\det g_6}\sqrt{1+e^{2\phi}\mathcal G^2_{\pm}}\mp \int_{W^6_{\pm}}\left(B_6-\mathcal G_{\pm} C_6\right)\right\}\ ,
\end{equation}
and $g_6$ is the induced metric on the worldvolume $W^6_{\pm}$ of the corresponding brane. The fields $\mathcal G_{\pm}$ are worldvolume scalars taking the following values
\begin{equation}
\mathcal{G_{\pm}}=\pm \frac{p}{2}-C_0=\pm \frac{p}{2}- M\Psi_{\pm}\ .
\end{equation}
The origin of the terms supported by $\mathcal{G}_{\pm}$ is explained in reference \cite{Gautason:2015tla} and can be understood as originating from a series of T-dualities from the NS5 action used by KPV \cite{Kachru:2002gs}. As such, we have the same issue here as in KPV where our dynamics is strongly coupled, hence we can only hope to derive features that are qualitatively reliable. We can motivate this further by referring to the recent computations for the polarisation potentials using the blackfold approach, but we postpone that to the discussion Section \ref{sec:discussion}.

The last term $S_{\text{int}}$ is an interaction term that takes into account both gravitational and Coulomb interactions. We will discuss it later.

Let us evaluate the action (\ref{eq:actionNS5}) on the background. Starting with  the DBI terms we get
\begin{equation}
\begin{aligned}
S^{\text{DBI}}_\pm=&-\frac{\mu_5 M}{2g_sv^2} \int \d^6x\, \sqrt{\left(\frac{p}{M}\mp 2\Psi_\pm\right)^2+\left(\f{2 v}{g_s M}\right)^2}\sqrt{-\det\left(\eta_{\mu\nu}+v^2\partial_\mu \Psi_\pm \partial_\nu \Psi_\pm\right)}\\
=&-\frac{\mu_5 M}{2g_sv^2} \int \d^6x\,\sqrt{\left(\frac{p}{M}\mp 2\Psi_\pm\right)^2+\left(\f{2 v}{g_s M}\right)^2}\,\left[1+\frac{v^2}{2}\left(\partial \Psi_{\pm}\right)^2+\dots\right]
\ .
\end{aligned}
\end{equation}
Notice the explicit appearance of the parameter $v$ which is a remnant of the function $S$ defining the flux background evaluated at the origin $r=0$. Here, the ellipsis stand for higher-derivative contributions that we shall ignore from now on.

The evaluation of the Wess-Zumino terms yield
\begin{equation}\label{eq:WZterms}
S^{\text{WZ}}_++S^{\text{WZ}}_-=- \mu_5 \int_{W^6}\left(\mathcal G_+-\mathcal G_-\right)C_6= -\mu_5M \int_{W^6}\left(\frac{p}{M} - (\Psi_+-\Psi_-)\right)C_6 \ .
\end{equation}
In the above expression it is manifest that the effective D5-brane charge is a function of the relative distance $\Delta \Psi=\Psi_+-\Psi_-$  between the pair, being $-p$ the initial charge $\Delta \Psi=0$ and $M-p$  when they meet together at the other side $\Delta\Psi=1$. Since the 6-form RR potential is given by
\begin{equation}
C_6=\frac{1}{g_sS}\,\d^6x\ ,
\end{equation}
we can simplify the Wess-Zumino action as follows:
\begin{equation}
S^{\text{WZ}}_++S^{\text{WZ}}_-=-\frac{\mu_5M}{g_s v^2}\int \d^6x\,   \left(\frac{p}{M}-(\Psi_+-\Psi_-)\right)\ .
\end{equation}

It is well-known that in the description of brane-antibrane dynamics, tachyon condensation \cite{Sen:1998sm} plays a crucial role when the branes are sufficiently close, as it is our case here. In reference \cite{Danielsson:2016cit} this was ignored and as we will explain now, the tachyon dynamics is crucial in getting a complete picture.

The $\overline{\text{NS5}}$-NS$5$-brane pair we are considering is not a pair of pure NS$5$-branes but they carry $\overline{\text{D5}}$-brane charges and so we should recover the action for $p$ $\overline{\text{D5}}$-branes when $\Delta\Psi=0$, where the branes are coincident. Instead, what we have is
\begin{equation}
S_++S_-=-\mu_5p\left\{\frac{2}{g_s^2 v p}\int \d^6x \sqrt{1+\frac{g_s^2p^2}{4v^2}}\sqrt{-\det \eta_{\mu\nu}}+\int_{W_6} C_6\right\}\ ,
\end{equation}
which does not vanish for $p=0$. This is because of the ``1'' in the square root. We can cure this by taking into account the dynamics of the tachyon field \cite{Sen:1998sm,Hashimoto:2002xt}, which effectively allows us to remove the ``1''.

Recall that in the case of a D-brane anti-D-brane pair in close proximity, the dynamics of the complex tachyon field $T$ is dictated by the potential:
\begin{equation}\label{eq:tachpot}
V\left(T\right) \approx \e^{-|T|^2}\left(1+L^2|T|^2\right)\, ,
\end{equation}
where $L$ is the distance between the brane and the anti-brane. The qualitative features of the tachyon potential depends crucially on the distance $L$, see Fig.~\ref{fig:tachyon}. We have two situations:
\begin{itemize}
\item $L<1$. This corresponds to the situation where the branes are close enough to annihilate each other into closed string radiation. The potential is a runaway towards $|T|\rightarrow \infty$, the true vacuum where $V \rightarrow 0$.
\item $L>1$. This in turn corresponds to the case when the branes are far enough so that the system has a meta-stable state at $T=0$ (which is not the true vacuum). Indeed, expanding the tachyon potential
\begin{equation}
V(T)=1+\left(L^2-1\right)|T|^2+\mathcal O\left(|T|^4\right)\ ,
\end{equation}
we can see that the mass of the tachyon field is positive in this limit.
\begin{figure}[ht!]
\begin{center}
\includegraphics{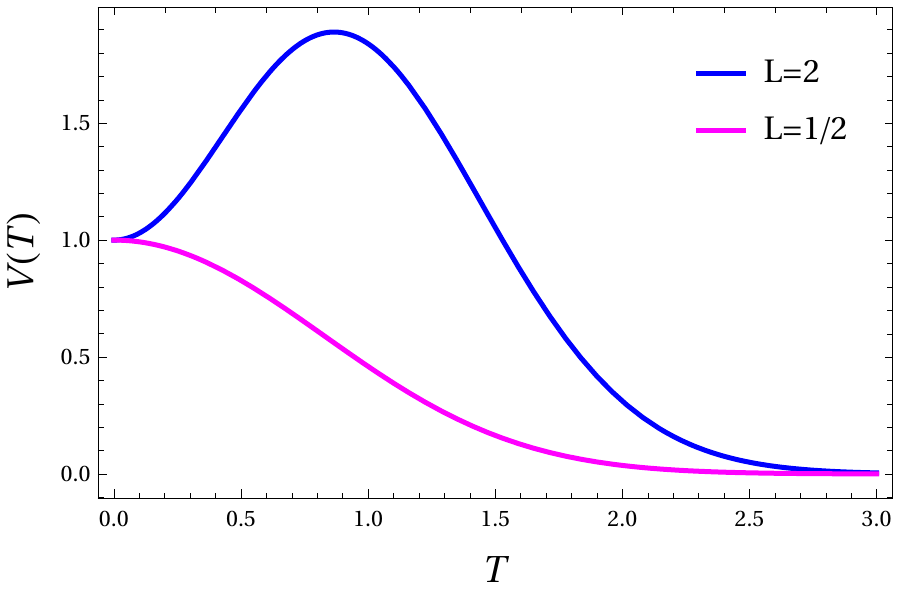}
\end{center}
\caption{\it The tachyon potential for two different values of the NS5/$\overline{\text{NS5}}$ distance.}
\label{fig:tachyon}
\end{figure}
\end{itemize}

Now we need to include this tachyon potential in our previous description of the dynamics of the NS5 dipole. Although tachyon condensation has only been studied in the literature for D-branes, we  expect on physical grounds\footnote{One could try to argue via S-duality but our main physical requirement is the simple condensation process: one should have zero energy when the branes are in the true vacuum and some non-zero energy is allowed when they are sufficiently separated. We therefore simply copy the D5 tachyon action and only change the dilaton coupling according to the way NS5 tension scales with the string coupling.} that the tachyon potential should replace the 1 in the square root appearing in the DBI term (\ref{eq:actionNS5}). The new form of the combined action is then
\begin{equation}
S_++S_-=-\mu_5p\left\{\frac{2}{g_s^2 v p}\int \d^6x \sqrt{V(T)+\frac{g_s^2p^2}{4v^2}}\sqrt{-\det \eta_{\mu\nu}}+\int_{W_6} C_6\right\}\,,
\end{equation}
where we have omitted the kinetic term for the tachyon as it does not play a role in our discussion.
The parameter $L$ in the potential then can be traded for the physical distance between the pair of NS5-branes. Note that when $L<1$ the tachyon potential is runaway and is therefore the tachyon remains in the true vacuum $|T|\rightarrow \infty$. This in turn implies that the tachyon potential vanishes $V(T)\rightarrow 0$. This is certainly the case for probe D5-branes which polarise to a pair of NS5-branes that are not separated by a large distance. As the NS5-brane pair moves away from each other it can happen that the tachyon potential develops a meta-stable minimum, but that is not relevant for our situation since the tachyon can consistently remain in the true vacuum for all values of $\Delta\Psi$. We therefore conclude that the effect of the tachyon is simple, it just removes the ``1'' in the square root.  This is the crucial observation that was missing in reference \cite{Danielsson:2016cit}. It at the same time justifies the computation carried out in \cite{Danielsson:2016cit} in which the ``1'' in the square root was simply ignored.\footnote{For D-branes polarising into larger branes, meta-stable states always form in a regime where the first term in the square root can be ignored. However, if one checks the details of the meta-stable state of \cite{Danielsson:2016cit} it exactly appears where that term could not have been ignored. This was left unnoticed in \cite{Danielsson:2016cit} and a main point of this paper is to show that when the tachyon is included one obtains a justification for putting that term to zero: it is simply because the tachyon remains in the true vacuum throughout the polarisation.}

As already mentioned earlier, the NS$5$-brane actions we are using is derived using type IIB S-duality and is therefore valid at strong coupling. We are on the other hand interested in the polarization potential at weak coupling. In a similar manner, the tachyon dynamics described here is not expected to be under control, and we can only possibly expect for qualitative features to remain true.  For the remainder of the computation we do not rely on any detail of the tachyon effective action aside from the qualitative feature that at small enough $\Delta \Psi$ the tachyon is in its true vacuum such that the first term in the DBI square root vanishes. As a consequence we recover the action for $p$ $\overline{\text{D5}}$-branes when the $\overline{\text{NS5}}$-NS$5$-brane pair coincides.

So with the tachyon in its true vacuum, the full effective potential for $\Delta\Psi$ is\footnote{The effective potential also depends on the combination $\Phi\equiv\Psi_++\Psi_-$. However, we have set $\Phi=0$ in Eq. (\ref{eq:eff-potential}), which is a valid simplification for our purposes since the meta-stable state is always there.}
\begin{equation}\label{eq:eff-potential}
\begin{aligned}
V_{\text{eff}}\left(\Delta\Psi\right)=\,&\frac{\mu_5 M}{g_sv^2 }\left\{\Bigg|\frac{p}{M} - \Delta \Psi\Bigg|+\frac{p}{M} - \Delta\Psi + V_{\text{int}}(\Delta\Psi) \right\}\ .
\end{aligned}
\end{equation}
where we included an interaction potential that corresponds to the energy required to separate the NS5-$\overline{\text{NS5}}$ pair, while being in the true tachyon vacuum. In \cite{Danielsson:2016cit} this interaction term was computed assuming the gravitational force between a NS5-$\overline{\text{NS5}}$ in flat space. But this assumes one is in the meta-stable tachyon vacuum, which is not relevant here. Luckily \emph{we only need a non-zero force that is not too big} in order to find a meta-stable state at
\begin{equation}
\Delta \Psi=p/M \,,
\end{equation}
as long as $p/M<1/2$.  The crucial aspect is that this meta-stable state carries exactly zero D5-charge since:
\begin{equation}
Q_5 = -\mu_5\big(p - M\Delta\Psi\big)\,,
\end{equation}
which can be read off from e.g.~Eq.~(\ref{eq:WZterms}). This meta-stable state exists as long as the interaction potential is ``not too big'', that is
\begin{equation}\label{toobig}
\partial_{\Delta \Psi} V_{\text{int}}(\Delta\Psi)<2.
\end{equation}
We want to emphasize once more that the actual details of the interaction potential are completely irrelevant to arrive at the sharp prediction $Q=0$. The only thing we require is a force between the NS5-$\overline{\text{NS5}}$ pair that is not overly big, meaning it satisfies (\ref{toobig}). We have explicitly checked that using the classical interaction potential, assuming no tachyon condensation we can even satisfy this inequality easily. Clearly that is a major overestimation since it does not take into account the way the NS5 tension vanishes towards the true tachyon vacuum. The only possible source of worry is that there is no interaction potential at all. We have no explicit computation to show there must be a potential, but on physical grounds it would be highly surprising that one could separate the NS5-$\overline{\text{NS5}}$ pair without some energy cost.

We conclude that the most interesting fact of our whole set-up is that \emph{we only need some non-zero force, whose details are irrelevant to arrive at the fact that the meta-stable state lives at exactly $Q_6=0$ (or better $Q_5=0$ in our T-dual framework).} This very robust feature of our set-up is what makes us confident about the $Q_6=0$ prediction.

There are a number of assumptions that go into this calculation, which we have commented upon already. We do however go over them yet again to make it clear where there is room for improvement. One of the assumptions that go in to our calculation is that we can apply the intuition regarding the tachyon dynamics borrowed from $\overline{\text{D5}}$-D$5$-brane pairs to our $\overline{\text{NS5}}$-NS$5$-brane pair. We find this trustworthy as it reproduce the original D5-brane as the $\overline{\text{NS5}}$-NS$5$-brane pair annihilate. Another approximation we use that is inherited from this dynamics, and that is that the tachyon remains in the true vacuum in the meta-stable minima. The justification for this comes from that there is nothing preventing us from tuning the ratio $p/M$ small, which sets $L$ small in (\ref{eq:tachpot}), which leads to the true vacuum. In the final expression for the effective potential, we also have to assume condition (\ref{toobig}) on the interaction potential. For the whole of this calculation, we have used worldvolume couplings that are induced by (the T-dual of) an S-duality, in a similar manner to that of KPV, which means the worldvolume is strongly coupled. This strategy has been shown to reproduce identical results from the point of view of supergravity, using the blackfold approach, see Section \ref{sec:discussion}, where the calculation is weakly coupled. The final test to whether our result is reliable would be to establish our meta-stable state in supergravity, which we approach in the next section.

\section{Remarks on regular solutions}\label{sec:regular}
We have argued that the no-go theorems against the existence of backreacted solutions with regular $H_3$-flux are evaded in a peculiar way predicted by our probe computation because  the polarised state has exactly zero anti-brane charge.

In this section we discuss numerical solutions to the general Ansatz (\ref{Ansatz}) which we argue to describe the polarised KK5 state at $r=0$. Before doing so we must discuss the UV singularity of the flux throat solution which we briefly mentioned below Eq.~(\ref{S-solution}). In Appendix \ref{app:O6} we argue that the singularity can be interpreted as the backreaction of an O6 plane sitting in the UV. This interpretation implies that proper quantisation of charges pushes the solution out of the supergravity regime. As a proper supergravity solution the O6 plane charge must take non-standard value such that fluxes can be large and the solution is under control. This is probably not consistent microscopically since orientifolds cannot be stacked.

Alternatively the background should be interpreted as a non-compact solution with a singular UV. This is likely the most fruitful interpretation in light of the numerical solutions presented below. The reason is simple, but worth spending some time on. Recall that without the anti-branes the compact flux solution is of the ``no-scale type'', by which we mean that a particular combination of dilaton and volume is massless. This can be traced back to the well-known no-scale properties of IIB compactifications to four-dimensional Minkowski space with 3-form fluxes \cite{Giddings:2001yu,Dasgupta:1999ss}, to which our solutions are T-duals of \cite{Blaback:2012mu, Blaback:2013taa}. This means that once we add the anti-branes, without the extra ingredients to stabilise the moduli, we expect an effective potential in seven dimensions that is runaway in what used to be the modulus direction. In compactifications to four dimensions this is remedied by the KKLT \cite{Kachru:2003aw} or Large Volume Scenario stabilisation method \cite{Balasubramanian:2005zx}. Note that this was not an issue encountered in the KPV construction of $\overline{\text{D3}}$-branes in the Klebanov-Strassler throat. This is because the KS throat is non-compact and the energy of the $\overline{\text{D3}}$ can escape to infinity. In the compact case the energy has nowhere to escape to and so it curves the four-dimensional space, or rather, it leads to time-dependence. Since we are not aware of a moduli-stabilising ingredient in seven dimensions à la KKLT we stick to the non-compact scenario of KPV. This means, similar to $\overline{\text{D3}}$-branes in the KS throat, that supersymmetry breaking with $\overline{\text{D6}}$-branes does not curve the 7D worldvolume but instead the energy of the $\overline{\text{D6}}$ is pushed into the UV singularity.

We conclude that the singularity in which the flux throat ``ends'' is not per se unphysical. One sign of that is the fact that it can be seen as an orientifold singularity when the numbers $K$ and $M$ take specific values and that for some of those values the whole flux throat can be seen as compact. However for the sake of studying meta-stable $\overline{\text{D6}}$ supersymmetry breaking it is necessary to not have a compact throat. This means that one should not consider the singularity as the end of the throat.

\subsection{Numerical solutions}
A probe computation has revealed that $\overline{\text{D6}}$-branes polarise into a KK5 dipole that carries exactly zero ($\overline{\text{D6}}$) charge. This is quite remarkable and seems a way to evade the no-go theorem of \cite{Blaback:2011nz} for solutions with infinite flux-clumping. However, we can say more. In the appendix of reference \cite{Blaback:2011pn} it was found that, of all possible (singular and regular) boundary conditions at $r=0$ only two options exist that can lead to well behaved solutions:
\begin{enumerate}
	\item Solutions with actual D6-branes at $r=0$.
	\item Solutions without any localised tension or localised (anti-) D6 charge.
\end{enumerate}
Clearly the first option is not what we want, since it would rather describe the end-point of brane-flux decay and not a meta-stable state. Option 2 naively seems to correspond to the empty flux throat, prior to the insertion of anti-branes. But that is not the only possibility. In fact it is exactly the boundary condition expected for our meta-stable system. We already argued that the charge has to vanish exactly. The absence of any localised tension is also predicted by the probe computation since it is a property of KK monopoles, they are locally smooth solutions without associated delta-functions that provide tension. Normally KK monopoles come with off-diagonal metric terms that cannot be gauged away, but we are describing KK5 \emph{dipoles} that are parallel to each other and pointlike in the 3D space. Hence the off-diagonal terms could arguably be gauged away.

 We now proceed with presenting numerical evidence that shows the presence of non-trivial solutions with the second boundary condition that passes non-trivial tests for being the meta-stable state found in the probe.

For the sake of numerics it is useful to rewrite the general Ansatz (\ref{Ansatz}) to the form of the Ansatz in Eq.~(\ref{eq:fluxthroat1}) for the empty throat solution (in string frame)
\begin{equation}\label{eq:nBPSansatz}
\begin{aligned}
\d s^2=&\, S_a^{-1/2}\d s_7^2+S_b^{1/2}\left(\d r^2+r^2\d\Omega_2^2\right)\,,\quad e^{\phi}=g_s\, S_f^{-3/4}\,, \\
F_2=&-g_s^{-1}\left(\frac{S_a^7 S_b}{S_l^8}\right)^{1/4}\star_3 \d S_l\,,\quad H_3= - g_sM \left( \frac{S_a^7 S_b^3}{S_f^6 S_l^4} \right)^{1/4} (1 - S_l^{(0)} S_l) \, r^2\,\d r\wedge \Omega_2\,.
\end{aligned}
\end{equation}
where we have also reinstituted the integration constant $S_l^{(0)}$, called $\alpha_0$ in e.g.~\cite{Blaback:2011nz}, that corresponds to a constant shift of the $C_7$ potential. When all functions $S_x := \{S_a, S_b, S_f, S_l\}$ are identified the expressions reduce to Eq.~(\ref{eq:fluxthroat1}). The equations of motion for this Ansatz provide five independent equations: The $F_2$ Bianchi identity, the dilaton equation of motion, and Einstein's equations along the diagonal components of the world-volume, $r$, and angular directions. From the result of \cite{Blaback:2011pn},\footnote{The result of \cite{Blaback:2011pn} is based on an analysis of the boundary conditions where the space is warped AdS$_7 \times S^3$, with fixed curvature. The boundary condition we use here is one that is compatible with both the AdS$_7 \times S^3$ and $\mathbb{R}^{6,1} \times \mathbb{R}^3$ warped Ansätze that we use here. We also use a slight misnomer here as we refer to the set $S_x$ and $S_x'$ at a fixed $r$ as \emph{boundary conditions} when the appropriate name would be \emph{initial} conditions.} we have that the boundary condition is given by
\begin{equation}
	S_x|_{r=0} = \textrm{Constants}\,,\quad S_x'|_{r=0} = 0\,.
\end{equation}
Numerically this boundary condition is not useful because the equations of motion are singular at $r=0$. This means that we must instead obey the above approximately at some point $r=\epsilon$ for some small $\epsilon$.\footnote{A singular behaviour of the form $q/r$ to any of the $S_x$ would create a deviation from $S_x'|_{r=\epsilon} \approx 0$ at the size of at most $q/\epsilon^2 \sim |S_x'|_{r=\epsilon}|$. Meaning for example having $\epsilon = 10^{-6}$ and $|S_x'|_{r=\epsilon}| = 10^{-7}$ leaves $q \lesssim 10^{-19}$ making singular terms irrelevant, and ensures that we will only be seeing the regular solution.} Such a choice of approximate boundary conditions is furthermore restricted by the equations of motion. Since we have four functions and five equations of motion, one equation can be written as an algebraic equation in $S_x$ and $S_x'$, which therefore impose a constraint on the boundary conditions we can choose. By a naive counting we then have three degrees of freedom: Three of the $S_x$ can be fixed by hand, while the last one is determined by this algebraic constraint, and $S_x'$ are chosen to be numerically small. In fact, these boundary conditions imply that the force of a probe D$6$-brane is zero in the IR, so we will also minimise the force there to keep the $S_x'$ small since it is not enough to keep the algebraic equation solved. We will return to mention some details about the implementation of this shortly.

A numerical solution obtained from solving this system of ordinary differential equations has to meet some criteria. First of all a solution is unique given a set of boundary conditions, but we have an algebraic constraint that not only have to be solved at the boundary as already mentioned, but throughout the full solution. Secondly, the solution we find should have a UV (large-$r$) behaviour that is the same as that for the BPS solution, Eq.~(\ref{S-solution}). This means that all $S_x$ should go to zero at the same point for large $r$. Thirdly we are looking for a solution that would display some polarised $\overline{\text{D}6}$-brane remnant at $r=0$ which we translate into a force towards $r=0$ for a probe D$6$-brane. We furthermore demand that this force vanishes towards the UV singularity since there the solution should look like the empty throat where D6 probes feel no force. Alternatively one could argue that the UV singularity is like an O$6$-plane, which exerts no force on a D6-brane. Our strategy is not to enforce any of these criteria in the solving in any way, but rather simply by varying the IR boundary conditions by hand.

We are able to produce solutions that obeys all of these criteria, and we present one such solution here. The solutions are derived using the Julia programming language \cite{Julia-2017}, in particular using the \texttt{DifferentialEquations.jl} package \cite{DifferentialEquations.jl-2017}. This solution is given by the boundary conditions
\begin{equation}
	S_x|_{r=\epsilon} \approx \left\{\begin{array}{l}
	0.984946   \\
  0.998187   \\
  1.01840    \\
  0.497034
	\end{array}\right.\,,\quad S_x'|_{r=\epsilon} \approx \left\{\begin{array}{l}
	-2.21457\cdot 10^{-8}\\
  -1.91789\cdot 10^{-7}\\
  \ \ \, 6.23101\cdot 10^{-8}\\
  -2.19209\cdot 10^{-8}
	\end{array}\right.\,.
\end{equation}
at $\epsilon = 10.0^{-6}$, for $m = 1.0$, $g_s = 0.5$, and $S_l^{(0)} = 0$. These boundary conditions are derived by choosing $S_l|_{r=\epsilon} \approx 0.6$ by hand, and all other boundary conditions to the BPS values at $r=\epsilon$, as a starting value for a local search that solves the algebraic equation, as well as minimises the zero-force condition. The local search is performed by the \texttt{Optim.jl} package \cite{mogensen2018optim}. The detailed derivation of the equations of motion, and the implementation that results in this solution can be found in \cite{notebook}.

\begin{figure}
	\begin{center}
		\includegraphics[scale=0.5]{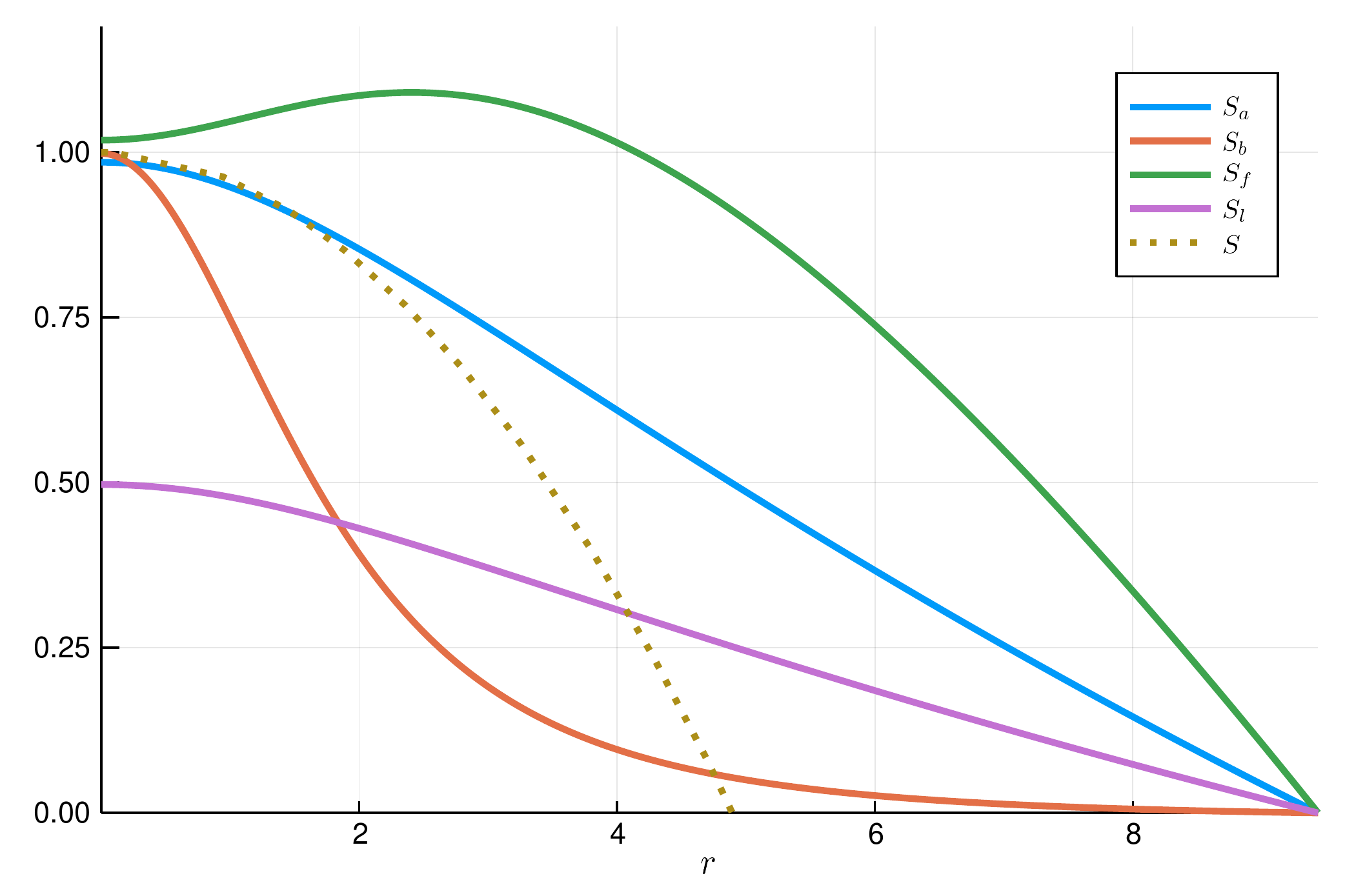}
	\end{center}
	\caption{\it The solution profiles for all functions, and a comparison to the analytical BPS solution $S$.}
	\label{fig:NBPSsol}
\end{figure}

\begin{figure}
	\begin{center}
		\begin{tabular}{cc}
			\includegraphics[scale=0.33]{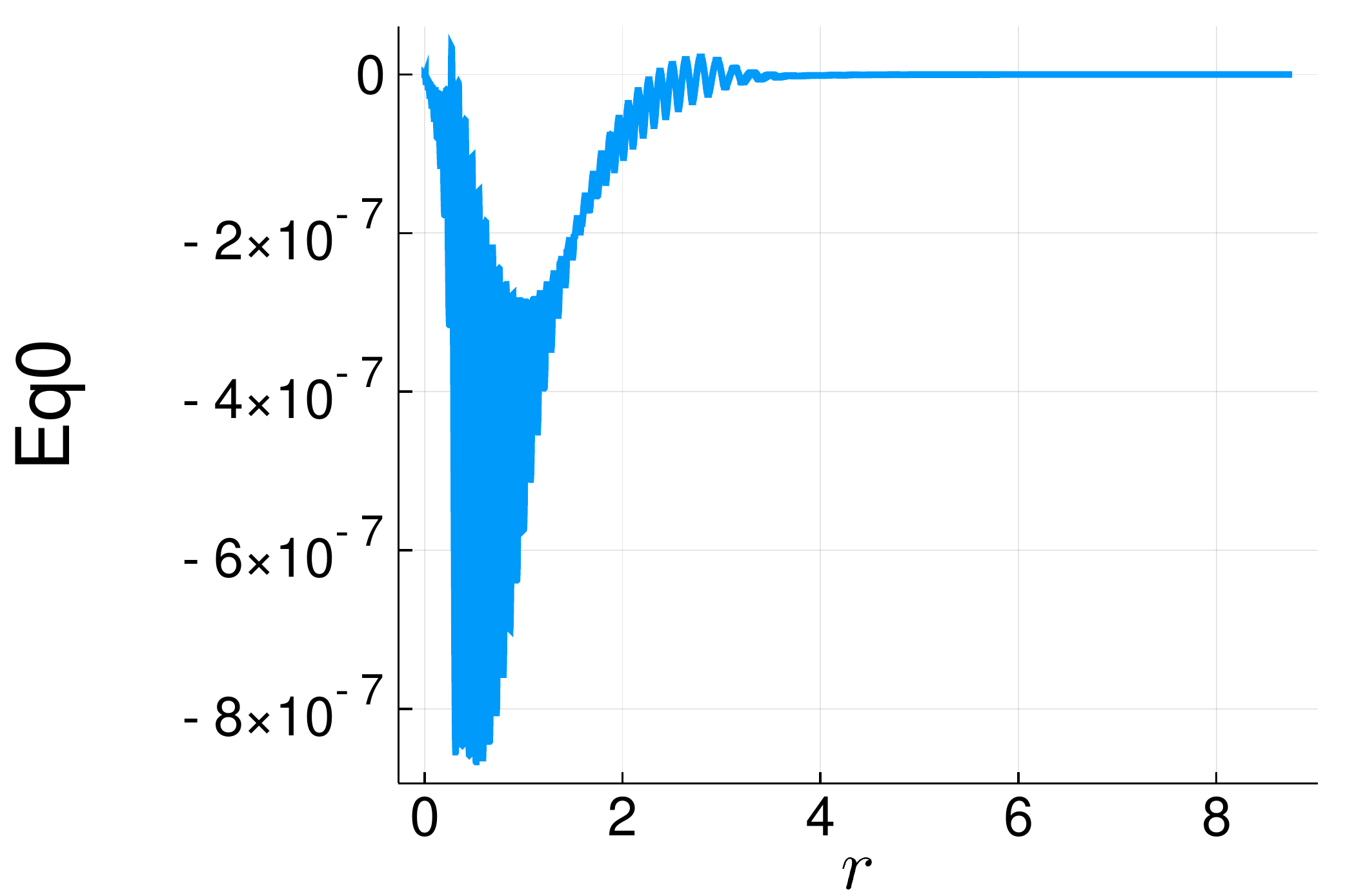} &
			\includegraphics[scale=0.33]{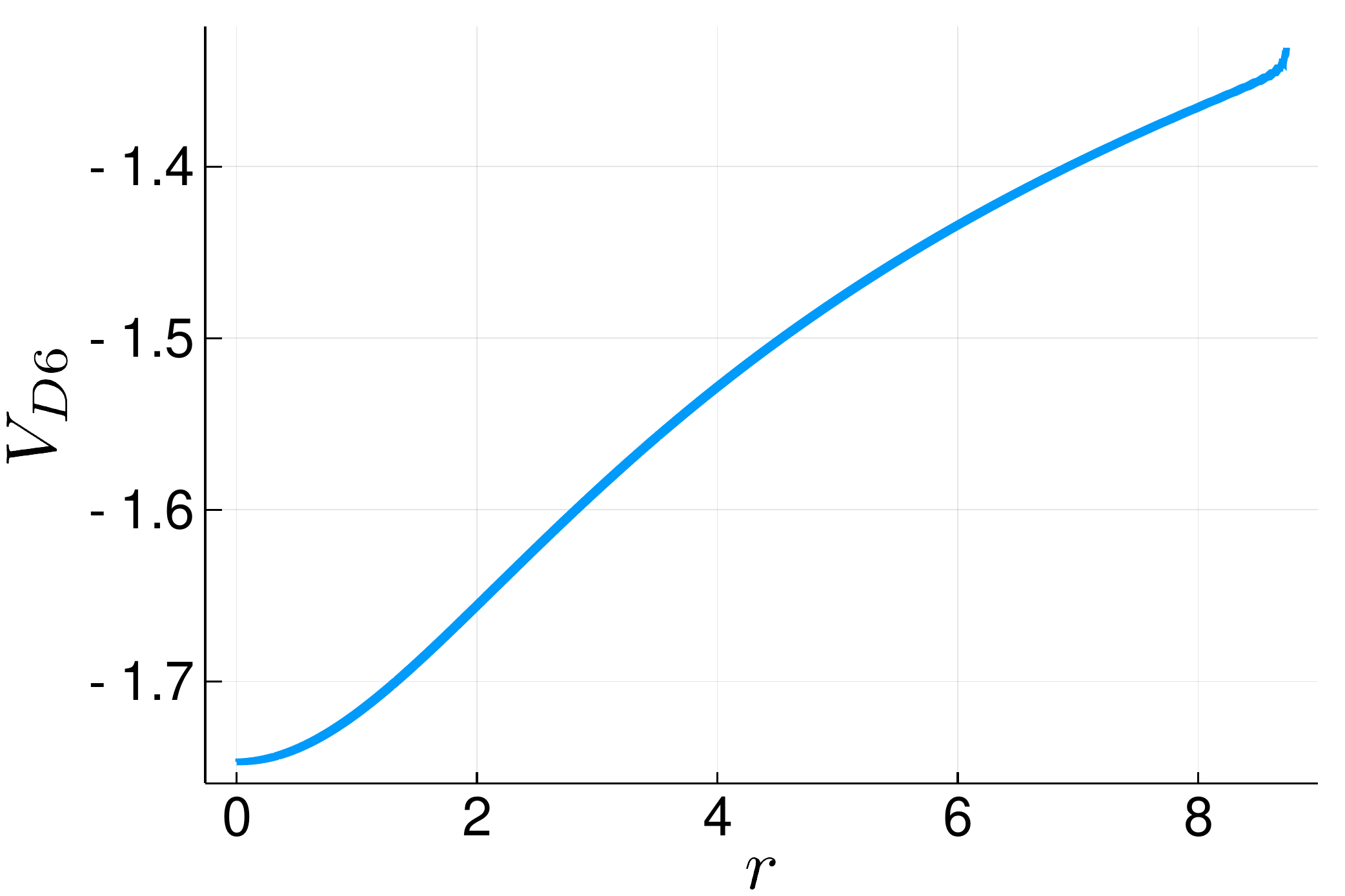} \\
			\includegraphics[scale=0.33]{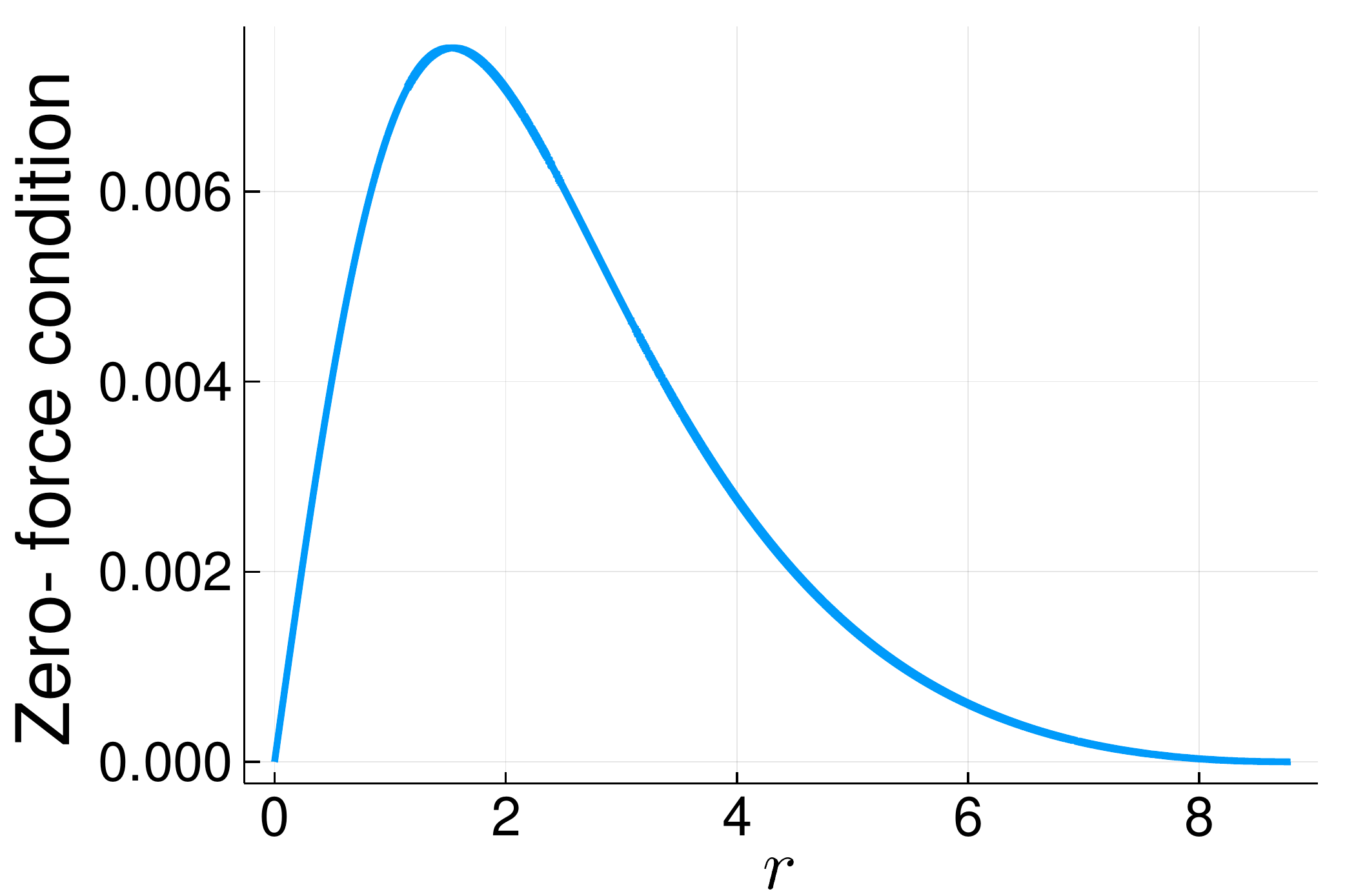} &
			\includegraphics[scale=0.33]{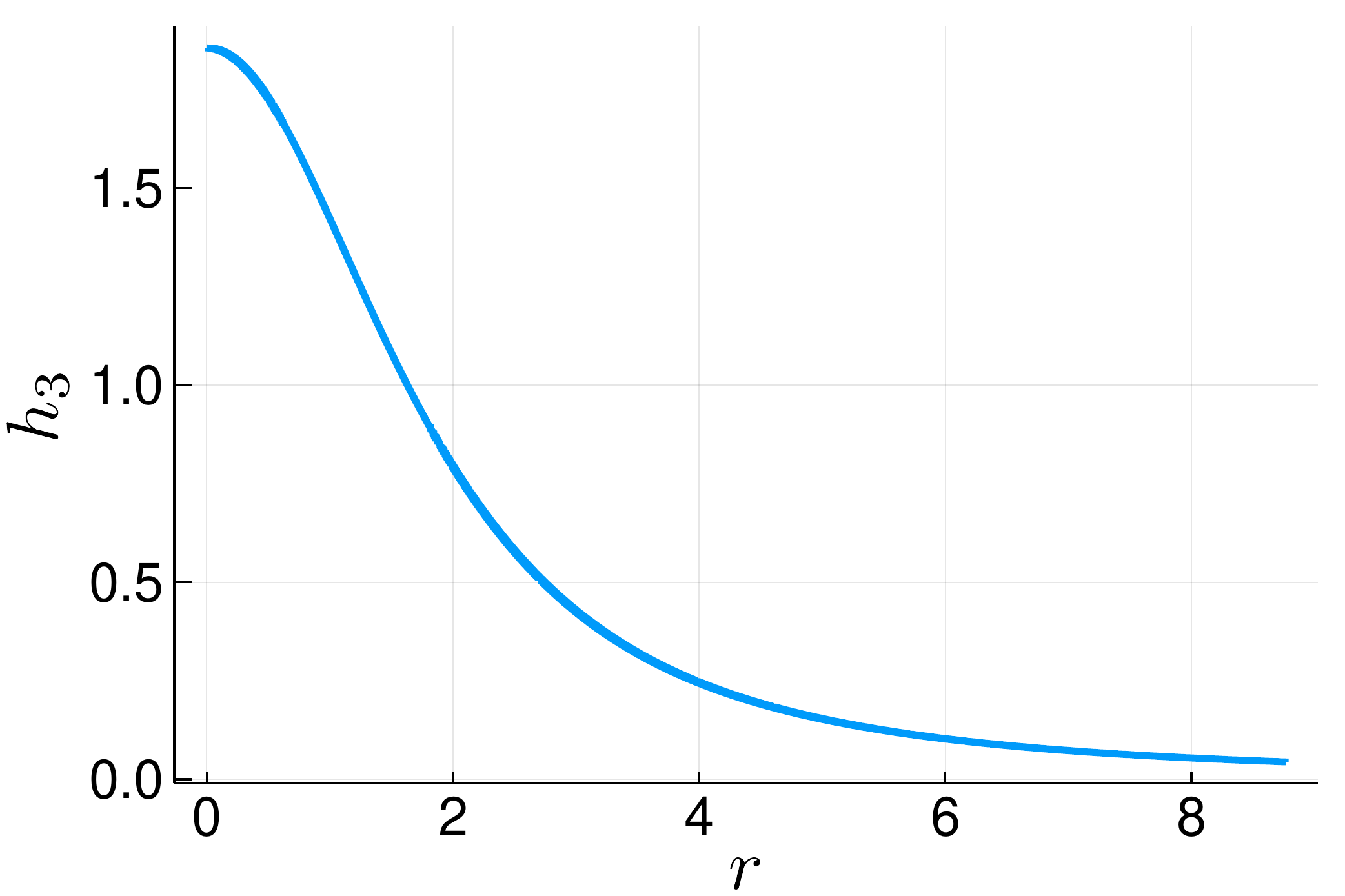}
		\end{tabular}
	\end{center}
	\caption{\it The upper left picture shows the deviation from zero of the algebraic equation. The upper right picture is a plot of the probe D$6$-brane potential. The lower left picture is of a zero-force condition, a condition that is such that it's zeroes are equivalent to zero force of a probe D$6$-brane, up to $1/S_x$-like singularities. The lower right picture is a plot of the $S_x$ dependence of $H_3$, hence showing flux clumping.	Because of the singular behaviour of the differential equations, or plotted functions, we have omitted the last few solutions points for the brane potential.}
	\label{fig:crit}
\end{figure}

The function profiles can be found in Figure \ref{fig:NBPSsol}. From this figure we can see that all of the functions terminate at $S_x = 0$ for large $r$. To demonstrate that this solution satisfies the criteria we set out previously, we also present Figure \ref{fig:crit}. There we see that the algebraic equation is fairly well respected throughout the solution, and we also have a probe D$6$-brane potential that slopes towards small $r$ as it should. The force expression is singular in the UV, so we have in Figure \ref{fig:crit} instead chosen to plot an expression that is regular in the UV but has the same zeroes as the force, that is, up to $1/S_x$ divergences. The plotted expression goes to zero at the UV, however some more analytical control is needed here to establish the zero force more firmly, as it is numerically plagued by $1/S_x$ singularities to various powers. In the same figure we also show the flux-clumping of $H_3$, which in the BPS solution would be identically $1$ for all $r$, but here we can see that the flux has clumped around $r=0$ and drained from the UV, and has a regular profile.

We find that this solution is consistent with the expectations of a non-BPS solution corresponding to the supergravity solution of a polarised $\overline{\text{D}6}$-brane remnant at $r=0$, and that it is numerically reliable. Certainly it would be interesting to see this realised analytically, even if only as a perturbation in the UV. We however leave this for future work.

\section{Discussion}\label{sec:discussion}

We have revised the results of \cite{Danielsson:2016cit} regarding the meta-stability of $\overline{\text{D6}}$-branes in massive IIA string theory. Our main result is that $\overline{\text{D6}}$-brane probes polarise into meta-stable KK5 bound states that carry  exactly zero $\overline{\text{D6}}$ charge. As a consequence the no-go theorems of \cite{Blaback:2011nz} against the existence of well-behaved backreacted supergravity solutions can be evaded and there is no obvious reason to doubt the meta-stability of anti-branes.

Furthermore, using the results in the appendix of reference \cite{Blaback:2011pn} we observe that a KK5 dipole without D6-brane charge is the unique supersymmetry-breaking source that can be introduced that does not imply an unphysical singularity. It is quite remarkable that a probe computation gives exactly that result. We have also presented numerical solutions with given sources as boundary conditions and they behave in the correct way in order to interpret them as the backreacted meta-stable states, although more detailed checks are probably possible and we plan to investigate this further in the future.

We believe that this closes a gap in the literature on anti-brane supersymmetry breaking. To our knowledge there is no scenario where singular fluxes could threaten the meta-stability of the probes, since this was already established for $\overline{\text{D3}}$-branes, $\overline{\text{M2}}$-branes and $\overline{\text{D}k}$-branes with $k<6$ in \cite{Cohen-Maldonado:2015ssa, Cohen-Maldonado:2016cjh} and now we established this for $k=6$.

However, some caution is required. The main shortcoming in our paper is a proper derivation of the brane polarisation potential. This issue plagues almost all probe computations that are used in understanding brane-flux decay and the key problem is the strongly coupled nature of the probe action.  In fact, if one wants to work in the supergravity regime, like we do, where $g_sp \gg 1$ there is no true ``probe approximation". Instead a different formalism has been developed known as the \emph{blackfold} approach \cite{Emparan:2009cs, Emparan:2009at, Armas:2016mes}. This approach is perfectly suited for understanding the dynamics of branes in the regimes we are looking at, since by construction it applies to the backreacting supergravity regime. Rather strikingly it was shown for $\overline{\text{D3}}$-branes in \cite{Armas:2018rsy} and for $\overline{\text{M2}}$-branes in \cite{Armas:2019asf} that the blackfold approach gives the identical critical points (and potential) as computed by the classical, yet strongly coupled, probe actions. An extra important virtue of the blackfold approach is that it allows to study the dynamics in presence of blackening factors. The results of \cite{Armas:2018rsy, Armas:2019asf} imply, amongst others, that anti-brane meta-stable states survive only a finite amount of temperature before getting destabilised, as expected on general grounds. We suspect that similarly the blackfold approach applied to $\overline{\text{D6}}$-branes will lead to identical results as the probe computations carried out in this paper and this is probably how ``supergravity knows" to evade any clash between probes and singularities. Together with our numerical analysis we therefore feel encouraged to trust our heuristic motivation of the probe action.

For the purpose of constructing  holographic duals to dynamical supersymmetry breaking, the supergravity regime is particularly relevant. The non-trivial AdS/CFT checks in \cite{DeWolfe:2008zy, Dymarsky:2013tna, Bertolini:2015hua, Krishnan:2018udc} (which partly rely on \cite{Bena:2009xk})  so far confirm that anti-brane meta-stable states are indeed dual to dynamical supersymmetry breaking\footnote{The interesting patterns of meta-stable states with thermalised probes observed in \cite{Armas:2018rsy, Armas:2019asf} so far have not been understood in the AdS/CFT context.
}. However the issues raised in  \cite{Bena:2014bxa, Bena:2014jaa, Bena:2015kia, Bena:2016fqp, Danielsson:2015eqa} should still be understood completely.

At first sight it seems that for the sake of cosmological model building a supergravity treatment of backreaction should not be of interest since the supergravity regime (large $g_sp$) is not relevant in that context. Certainly this is the case for $\overline{\text{D3}}$-branes in the KKLT scenario \cite{Kachru:2003aw} where one has to rely on the arguments of \cite{Michel:2014lva} that apply in the stringy regime. Concerning classical dS solutions with O6 planes in massive IIA it was argued in \cite{Kallosh:2018nrk} that $\overline{\text{D6}}$-branes could be the crucial ingredients to evade the ubiquitous tachyons in these scenarios \cite{Danielsson:2012et,Junghans:2016abx}. Interestingly, relying on the results of \cite{Banlaki:2018ayh,Junghans:2018gdb} one can infer that the only obvious way to evade strong coupling problems of the flux background is to cook up a geometry with a large amount of O6 planes, which in turn requires a large amount of $\overline{\text{D6}}$-branes. This forces the $\overline{\text{D6}}$-branes into the classical supergravity regime. If geometries with large O6 plane charge are to be found in the future, then our findings imply that the problem of possible direct brane-flux decay, due to backreaction of the sources, is not a concern.

\section*{Acknowledgements}

We thank Ulf Danielsson for collaborations on an earlier related work, and Iosif Bena and Vasilis Niarchos for comments on the manuscript. We would like to thank Xin Gao for his input. We are also grateful to all the developers and contributors of Jupyter, \texttt{IJulia.jl}, \texttt{Plots.jl}, \texttt{ParameterizedFunctions.jl}, and \texttt{LaTeXStrings.jl} for Julia. The work of JB is supported by MIUR-PRIN contract 2015MP2CX4002 ``Non-perturbative aspects of gauge theories and strings''.  The work of AR is supported by a ``Centro de Excelencia Internacional UAM/CSIC'' FPI pre-doctoral grant. The work of TVR is supported by the KU Leuven C1 grant ZKD1118 C16/16/005 and the FWO odysseus grant G.0.E52.14N and by the C16/16/005 grant of the KULeuven.

\appendix
\section{A second look at the flux throat solution}\label{app:O6}
In this appendix we focus on the flux background solution given in  \eqref{eq:fluxthroat1}. As we noted the flux background is singular at a finite distance from the origin. We will demonstrate that the fields have the correct local behaviour such that the singularity can be interpreted as an O6-planes singularity.

Let us first rewrite the solution (\ref{S-solution}) in a more convenient way as follows:
\begin{equation}
S =   \frac{(Mg_s)^2(r_0^2-r^2)}{6} ~,
\end{equation}
The solution has a peculiar singularity for $r\to r_0$. In order to examine the local behaviour of the singularity we expand around it, by introducing a new local coordinate $x$:
\begin{equation}
	r = r_0 - \delta r\quad\text{with}\quad (\delta r)^5 =\f{6}{2 r_0 M^2 g_s^2}\left(\f{5}{4}x\right)^4~,
\end{equation}
and find
\begin{equation}\label{fluxsingularity}
	\d s^2\approx\left(\f{12}{5r_0M^2g_s^2  x}\right)^{2/5}\dd s_7^2 + \d x^2 + r_0^2\left(\f{5r_0M^2 g_s^2 x}{12}\right)^{2/5}\dd\Omega_2^2~.
\end{equation}
Let us compare this with an  O6-plane singularity. The metric of an O6 plane (in flat space) is given by
\begin{equation}
	\d s^2= \f{1}{\sqrt{h_\text{O6}}} \dd s_7^2 + \sqrt{h_\text{O6}}\left(\dd r^2 + r^2 \dd\Omega_2^2\right)~,
\end{equation}
where
\begin{equation}
	h_\text{O6} = 1-\f{Ng_s}{4\pi r}~,
\end{equation}
with $N$ the quantised O-plane charge (positive in our convention). When we expand this metric around the critical radius
\begin{equation}
	r = r_c + \delta r = \f{Ng_s}{4\pi}+ \delta r~.
\end{equation}
we find
\begin{equation}
	\d s^2 \approx \sqrt{\f{Ng_s }{4\pi \delta r}} \dd s_7^2 + \sqrt{\f{4\pi\delta r}{Ng_s }}\left(\dd \delta r^2 + \left(\f{Ng_s}{4\pi}\right)^2 \dd \Omega_2^2\right)~.
\end{equation}
By changing coordinates
\begin{equation}
	x \equiv \f45 \left(\f{4\pi}{Ng_s \ell_s}\right)^{1/4} \delta r^{5/4}~,
\end{equation}
we obtain the final expression
\begin{equation}
	\d s^2 \approx \left(\f{Ng_s }{5\pi x}\right)^{2/5}\dd s_7^2 +\d x^2 + \left(\f{Ng_s }{4 \pi}\right)^2\left(\f{5\pi x}{Ng_s }\right)^{2/5}\dd \Omega_2^2~.
\end{equation}
Here we see that the local structure of the singularity of our flux background in \eqref{fluxsingularity} has exactly the same local form as the near-singularity expansion of an O6-plane. The same holds true for the other supergravity fields with the exception of the NSNS three form $H_3$ and the Romans mass parameter. We would therefore like to interpret the flux background as a Minkowski compactification on a three-dimensional orientifold. This is similar in spirit to the GKP background which are four-dimensional compactifications of type IIB supergravity on Calabi-Yau orientifolds. In order for this interpretation to make sense we must ensure that fluxes can be quantised at the same time as cancelling the tadpole. The volume of the \emph{unwarped} manifold is just that of a ball of radius $r_0$:
\begin{equation}
	V = \f{4\pi r_0^3}{3}~.
\end{equation}
The warped volume is
\begin{equation}
	{\cal V} = \f{\pi^2 M g_s r_0^4}{4\sqrt{6}}~.
\end{equation}
Flux quantisation for the solution at hand this means $M\in {\bf Z}$ and
\begin{equation}
K\equiv\int H =M g_s V \in {\bf Z}~.
\end{equation}
The quantisation of $F_2$ is not required since the quantised charge is the Page charge which is the integral of $F_2-MB_2$ which is identically zero for our background. The orientifold charge itself must of course be that of a string theory O6-plane
which is
\begin{equation}\label{OplaneCharge}
2=N_{O6} \equiv \lim_{r\to r_0} \int F_2 =  M^2 g_sV  = M K~.
\end{equation}
Hence both $M$ and $K$ are bounded by 2. This does show that the tadpole condition is satisfied. A typical problem with stabilised backgrounds supported by O-planes is that they will not be in controllable regime of string coupling constants. This is also the case here. In order for our supergravity background to be trustworthy we must make sure the dilaton and curvature are small. Close to the regular point $r\to0$ these take the form
\begin{equation}
	\e^{2\phi}\big|_{r\to 0} \sim \f{1}{K M^2}~,\quad R\big|_{r\to 0}  \sim \f{1}{K}~.
\end{equation}
These can at best be 1/2 because of \eqref{OplaneCharge} which pushes the solution outside the regime of validity of supergravity because the numerical factors we have dropped are of order 10. The way around this problem is to consider O6 planes which carry a large D6-brane charge. These are of course not the proper O6-planes in string theory but the solutions may still be worthwhile to study.
One question that remains is what is the underlying manifold? A possible answer is that it is simply ${\bf R}^3$ with a ${\bf Z}_2$ action:
$r\mapsto \f{r_0^2}{r}~,$
that renders the manifold compact.

\bibliographystyle{utphys}

    \bibliography{refs}

\end{document}